\documentclass[aps,twocolumn,groupedaddress,showpacs]{revtex4}

\usepackage{natbib}
\usepackage{graphicx}
\begin{document}
\title{Experimental study of nonlinear focusing in a magneto-optical trap using a Z-scan technique }

\author{Yingxue Wang}
 \email{yingxuewang@wisc.edu}
\author{M. Saffman}
\affiliation{Department of Physics, University of Wisconsin-Madison, Madison, Wisconsin, 53706}

\date{\today}

\begin{abstract}
We present an experimental study of Z-scan measurements of the nonlinear response of cold Cs atoms in a magneto-optical trap. Numerical simulations of the Z-scan signal accounting for the nonuniform atomic density  agree with experimental results at large probe beam detunings. At small  probe detunings a spatially varying radiation force modifies the Z-scan signal. We also show that the measured scan is sensitive to the atomic polarization distribution. 
\end{abstract}
\maketitle

\section{Introduction}
\label{sec1}

Clouds of cold atomic vapors produced by laser cooling and trapping
techniques provide strongly nonlinear effects when probed with near resonant
light beams. Soon after the first demonstration of a magneto optical 
trap(MOT)\cite{chu86} it was recognized that the cold and dense clouds were
useful for studies of nonlinear
spectroscopy\cite{kimble_nls,grynberg1}, generation of non-classical
light\cite{giacobino} and other nonlinear effects\cite{Walker90,Tabosa03}.
In this work we study nonlinear focusing of a near resonant probe beam taking into account the spatial dependence of the atomic density and the effect of the probe induced radiative forces on the MOT distribution. 

From the
perspective of nonlinear optics a cold vapor in a MOT is an interesting
alternative to the nonlinearity of a hot vapor in a cell. A typical MOT
provides a peak density of order $10^{9}-10^{11}~\rm cm^{-3}$, and an
interaction length of a few mm. On the other hand a heated vapor cell can
easily have atomic  densities of $10^{14}~\rm cm^{-3}$ and interaction
lengths of 10 cm, or more, so that much stronger nonlinearities can be
achieved in a vapor cell than in a MOT.
Nonetheless there are a number of reasons for looking more closely at the
optical
nonlinearity provided by a cold vapor. To begin with the relative strength
of the MOT nonlinearity compared with a hot cell is larger than the above
estimates would indicate because of Doppler broadening effects. In the
presence of the large inhomogeneous Doppler width in a hot cell it is typical
to detune the probe beam by a GHz or more to avoid strong absorption.  Conversely the Doppler limited linewidth of cold atoms in a MOT is of order the
homogeneous linewidth or less so that much smaller detunings of order tens of
MHz can be used. The full saturated  nonlinearity can therefore be achieved
with a much weaker probe beam in a MOT than in a vapor cell. Additionally,
the large detunings used in vapor cells imply that detailed modeling of
nonlinear effects in Alkali vapors must account for the plethora of
hyperfine and Zeeman levels. Such  modeling, as  has been done by several
groups \cite{swiss,rubenstein}, involves numerical integration of hundreds of  coupled equations for the density matrix elements. Only in special cases,
using e.g. a buffer gas to create large pressure broadening, can a
simplified two-level type model provide an accurate description of nonlinear
propagation effects\cite{lange1}.  Working in a MOT with a small probe beam detuning the probe  interacts strongly
with only a single atomic transition so that the main effects of  nonlinear propagation can be achieved using a compact
two-level atomic model. As we show below this is only partially true  since  radiative forces and population distribution among Zeeman sublevels lead to observable features in the measured probe transmission.

In addition a cold atomic vapor potentially provides a qualitatively
different nonlinearity than a hot vapor does because the mechanical effects
of light can  result in strong modification of the atomic density
distribution, which in turn feeds back on the nonlinearity seen by the
optical beam. Indeed some experiments already observed reshaping of the MOT
cloud due to radiation trapping effects\cite{Walker90,Sesko91,Bagnato}.
While such density redistribution
effects in both position and momentum space may also occur in hot vapors, as in the collective atomic recoil
laser\cite{bonifacio1,lippi,bigelow}, and radiation pressure induced dispersion\cite{grimm_mlynek}, such effects  are potentially much more
pronounced in cold vapors where momentum transfer from the light beams is
significant. In particular we expect that complex spatial structures can be
formed due to coupled instabilities of the
light and density distributions in a fashion analogous to the effects that
have been predicted for light interacting with coherent matter
waves\cite{saffman98,saffmanskryabin}. Some recent related work has shown
evidence of nonlinear focusing in a MOT\cite{kaiser}, and possibly structure
formation in experiments that include the effects of cavity
feedback\cite{vuletic}.  In Sec. \ref{sec4} we discuss the
relevance of the present measurements in the context of observation of
coupled light and matter instabilities.

Our primary interest in the present  work is a detailed study of  nonlinear focusing and defocusing of a tightly focused probe
beam that propagates through a MOT. The Z-scan technique was originally
developed\cite{Bahae89} for characterization of thin samples
of nonlinear materials. Since the MOT cloud is localized to a region of a
few mm in thickness we can easily apply this technique for characterization
of the MOT nonlinearity. The theoretical framework based on a two-level
model is described in Sec. \ref{sec2}. Z-scan measurements were taken with a
Cs MOT using the procedures discussed in Sec. \ref{sec3}. The experimental
and theoretical results are compared in
Sec. \ref{sec4} where we also compare additional measurements and
calculations of  reshaping of the transverse profile of the probe beam after
propagation through the MOT.

\section{Numerical Model}
\label{sec2}
In Z-scan measurements, the transverse profile of a laser beam passing through a nonlinear sample is investigated. 
In the presence of self-focusing or self-defocusing the transmittance through a small aperture placed after the medium exhibits a S-shaped dependence on the position of the beam waist with respect to the nonlinear sample. Following the original demonstration of the Z-scan technique\cite{Bahae89,Bahae90} there have been a large number of theoretical studies of the Z-scan method that take into account different types of nonlinear response and consider different characteristic ratios between the Rayleigh length of the probe beam and the thickness of the nonlinear sample\cite{Bahae90, Hermann93, Bian99, Yao03, Tsigaridas03, Zang03}. The interaction of a probe beam with a cloud of cold two-level atoms is described by a saturable mixed absorptive and dispersive nonlinearity with a susceptibility of the form\cite{Boyd}
\begin{equation}
\label{eqchi}
\chi({\bf r})=n_{a}({\bf r})W_0\frac{3\lambda^3}{4\pi^2}\frac{2\Delta/\gamma-i}{1+4\Delta^2/\gamma^2+I/I_s}.
\label{eq.chi}
\end{equation}
Here $n_{a}({\bf r})$  is the density of atoms at position $\bf r$, $W_0$ is the population difference in thermal equilibrium,  $\gamma$ is the homogeneous linewidth, $\Delta=\omega-\omega_a$ is the difference between the probe frequency $\omega=2\pi c/\lambda$ and the atomic transition frequency $\omega_a,$ $\lambda$ is the wavelength of the laser beam in vacuum, $I_s$ is the on resonance saturation intensity, and $I$ is the optical intensity.

None of the existing theoretical treatments can be directly used for our situation. Although the work of Bian, et al.\cite{Bian99} studied the Z-scan behavior for a saturable nonlinearity, it was restricted to no absorption and weak saturation. In work to be published elsewhere 
we derive analytical expressions for the Z-scan curve for  a susceptibility of the type given in Eq. (\ref{eq.chi}).  However, we find that in order to obtain good agreement with experimental measurements it is necessary to take account of the spatial variation of the density in the MOT cloud. We have therefore relied on direct numerical simulations 
to compare with experimental results. To do so we assume a scalar probe beam of the form 
$ E=[A({\bf r})/2]e^{\imath(kz-\omega t)}+[A^*({\bf r})/2]e^{-\imath(kz-\omega t)}$
and invoke the paraxial and slowly varying envelope approximations to arrive at the wave equation
\begin{equation}
\label{eqPWEnorm}
\frac{\partial{A({\bf r})}}{\partial{z}}-\frac{i}{2k}{\nabla}_\perp^2A({\bf r})=i \frac{k}{2}
\chi({\bf r}) A({\bf r})
\end{equation}
where $\nabla_\perp^2=\partial^2/\partial x^2 + \partial^2/\partial y^2$
with $x,y$ the transverse coordinates and $k=2\pi/\lambda.$ 

Equation  (\ref{eqPWEnorm}) was solved numerically on a $128\times128$ or $256\times 256$ point transverse grid using a split-step spectral code. Propagation from the atomic cloud to the pinhole plane was calculated by 
solving  (\ref{eqPWEnorm}) with the right hand side equal to zero. As the experimental free propagation distance of 15 cm resulted in the beam spilling over the edges of the computational window, calculations were done with a distance of 3 cm, and a pinhole diameter of $1/5$ the actual size.  
The numerical solutions obtained in this way describe the interaction of the probe beam with an unperturbed MOT. Since there is no integration over the Doppler profile of the atoms we are implicitly assuming they are at rest. 
In reality the atoms are accelerated due to the absorption of photons from the beam, so the laser frequency seen by the atoms is shifted to the red.  A model including this Doppler effect should be used when the probe beam interacts with the cloud for a time corresponding to many absorption/emission cycles. We discuss the implications of the radiative force for the Z-scan curves in Sec. \ref{sec4}.

\section{Experimental Setup}
\label{sec3}

\begin{figure}[!t]
\includegraphics[width=8.4cm]{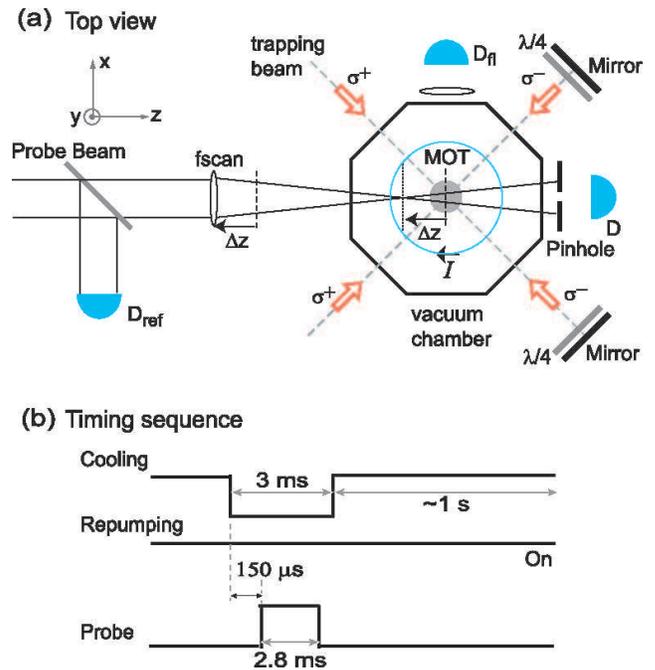}
\caption{(color on-line) Z-scan experimental setup (a) and timing sequence (b). D and D$_{ref}$ are signal and reference detectors respectively and D$_{fl}$ is the fluorescence detector used to monitor the number of cold atoms.}
\label{fig.ZscanSetup}
\end{figure}

The experimental setup is shown in Fig. \ref{fig.ZscanSetup}(a). A standard MOT was 
loaded directly from a background Cs vapor. The trapping and repumping beams were obtained from two external cavity diode laser systems. The trapping beams were detuned from the $F=4\to F'=5$ cycling transition by $\Delta\sim-2\gamma$, where $\gamma/(2\pi)=5.2$ MHz. 
The Gaussian diameter of each trapping beam was 2.5 cm with a peak intensity of 2.4 mW/cm$^2$, and the beams were retroreflected after passage through the MOT. The peak saturation parameter for all six beams was 13.1. The repumping beam was tuned to the $F=3\to F'=4$ transition. The magnetic field gradient was 9 G/cm along the vertical 
$y$ direction while the probe beam propagated horizontally along $z.$ Time of flight measurements gave a typical MOT temperature of 60 $\mu$K. 

\begin{figure}[!t]
  \includegraphics[width=8.6 cm]{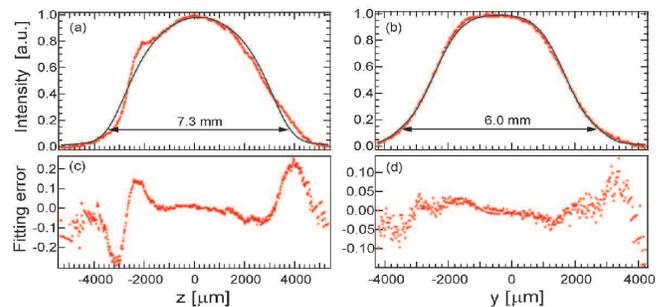}
\caption{(color on line) Intensity profiles (dots) of the Cs atomic cloud along $z$ (a) and $y$  (b) directions, with curve fitting results (solid lines).
Frames (c) and (d) show the corresponding fitting errors.}
\label{fig.Cloud}
\end{figure}

In order to model the Z-scan data accurately some care was taken to characterize in detail the spatial distribution of trapped atoms. 
Fluorescence measurements of the MOT cloud taken with a camera placed on the $x$ axis revealed a flattened density profile indicative of radiation trapping effects\cite{Walker90,Sesko91,Townsend95}. This type of profile has previously been modeled with a Fermi-Dirac type distribution\cite{Hoffmann94}. We chose to use an expansion with more fitting parameters  of the form
\begin{eqnarray}
n_a(y,z)&=&n_{a0}\frac{a_z\,e^{-\frac{2z^2}{w_z^2}}+b_z\,e^{-\frac{2z^4}{w_z^4}}+c_z\,e^{-\frac{2z^6}{w_z^6}}}{a_z+b_z+c_z}\nonumber \\
&&\times \frac{a_y\,e^{-\frac{2y^2}{w_y^2}}+b_y\,e^{-\frac{2y^4}{w_y^4}}+c_y\,e^{-\frac{2y^6}{w_y^6}}}{a_y+b_y+c_y}
\label{eqnatom}
\end{eqnarray}
where $n_{a0}$ is the peak density, $w_z$, $w_y$ are the Gaussian radii along $z$ and $y$ and the $a,b,c$ are fit parameters along the two axes. 
 The trapped Cs atoms formed an ellipsoidal cloud, which was modeled as a density distribution Eq. (\ref{eqnatom}) with typical Gaussian diameters of $2 w_z=7.3 ~\rm mm$ and $2 w_y=6.0 ~\rm mm$  as shown in Fig. \ref{fig.Cloud}.  The fit residuals, as seen in the lower plots in Fig. \ref{fig.Cloud}, were small in the center of the MOT and reached  20\% at the very edges of the cloud. Since the axis of the B field coils was along $y$ we assumed that the $x$ and $z$ density profiles were equal. As the size of the probe beam was small compared to the width of the cloud we approximated $n_a({\bf r})$ by $n_a(z)$ given by the first line of Eq. (\ref{eqnatom}) in our numerical simulations. 

The total number of trapped atoms was $2.5\times10^8$ as measured by an optical pumping method \cite{Chu92,Chen01}. This number was also measured and monitored from day to day by recording the fluorescence signal using a lens and calibrated photodetector \cite{Townsend95}. The result from the fluorescence signal, taking into account the corrections for the atomic polarizability distribution discussed in Ref. \cite{Townsend95} was about 2.5 times lower than that from the optical pumping measurement. The number of cold atoms measured by the fluorescence method  varied by  up to 15$\%$ from day to day. The peak atomic density determined from the number measurement using the optical pumping method, and the size of the cloud using the data shown in Fig. \ref{fig.Cloud}, was typically $5.5\times 10^{10}~\rm cm^{-3}$, and $2.2\times 10^{10}~\rm cm^{-3}$ using the fluorescence measurement of the number of atoms. We believe the optical pumping method to be more accurate since it does not rely on any assumptions about the polarization of the MOT cloud. 

The Z-scan probe beam was derived from the trapping laser and frequency shifted to the desired detuning with an acousto-optic modulator. 
The beam was then spatially filtered with an optical fiber before focusing into the MOT with a lens $f_{scan}=400~\rm mm$ mounted on a translation stage.  The Gaussian radius of the beam at the focus was $w_0=24.5 ~\mu\rm m.$ A 1.0 mm diameter pinhole was placed 15 cm away from the center of the MOT (outside the vacuum chamber). The transmitted field through the pinhole was measured by photodetector $D$. 
To measure the transmittance of the probe beam, the trapping beams and the probe beam were turned on sequentially, as shown in Fig. \ref{fig.ZscanSetup}(b).
During the measurement, the trapping beams were turned off first and the repumping beam was left on to pump the atoms into the  $F=4$ state. The transmittance of the pinhole at different times after the probe beam was turned on was measured as $D(t)/D_{\rm ref}.$

In the experiment, the lens was scanned instead of the nonlinear sample as shown in Fig. \ref{fig.ZscanSetup}(a). Compared with  traditional Z-scan measurements, there are two points to consider. First, note that a movement of the lens in the $+z$ direction is equivalent to 
a movement of the cloud in the $-z$ direction. Therefore, Z-scan curves obtained in this experiment have the opposite configuration to the traditional ones, e.g. a peak followed by a valley shows a self-focusing nonlinearity. Second, since the lens was scanned, the position of the beam waist changed, which affected the linear (without cold atoms) transmittance. To take this effect into account, we recorded $D^{(\rm on/off)}$ with the MOT on and 
off (by turning the magnetic field on and off) so that the normalized Z-scan curve was given by $D^{(\rm on)}/D^{(\rm off)}$
as a function of $z$.

\begin{figure}[!b]
\includegraphics[width=8.6 cm]{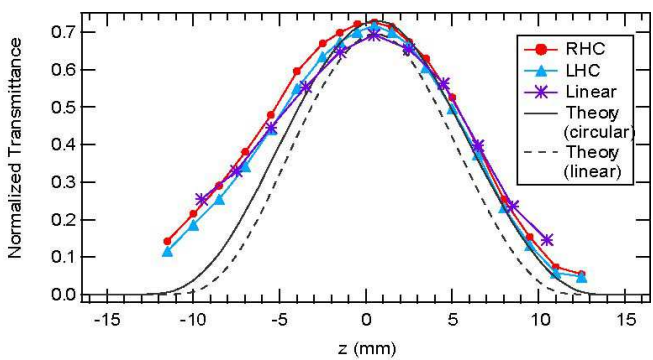}
\caption{(color on-line) Measured on resonance transmission  curve of a Cs cloud at $\lambda=852$ nm $6 \mu$s after the probe beam was turned on with different polarizations. The solid line is the calculated result for a circularly polarized beam and the dashed line is for a linearly polarized beam. The Gaussian diameter of the atomic cloud was $2w_z= 6.2~\rm  mm$ and $a_z=0.85,b_z=0.03,c_z=0.07$.}
\label{fig.Abs}
\end{figure}

Before discussing the Z-scan measurements we note that  transmission scans with the pinhole removed can be used for calibration of the peak atomic density. 
Figure \ref{fig.Abs} shows the measured transmittance curve with the pinhole removed and the probe beam tuned on resonance with the transition $F=4\to F'=5.$ The peak intensity was 1.07 W/cm$^2$
giving a peak saturation parameter of 975. Measurements were taken for both circular ($\sigma^+$ and $\sigma^-$) and linear polarized beams. The measurements show that the probe beam was strongly absorbed when the beam waist was far from the center of the atomic cloud. As the beam waist was moved closer to the cloud center the intensity of the beam increased and the absorption saturated, so that the transmittance increased. The results using right and left hand circular polarization were very similar to each other, while the curve for the linear beam is broader than those with circular beams and its peak transmittance is a little lower. The solid line is the numerical calculation under the experimental conditions using a two-level atom model with I$_s$ = 1.1 mW/cm$^2$,
and the dashed line is the calculation with I$_s$ = 1.6 mW/cm$^2.$ Thus the solid line corresponds to a circularly polarized probe with the assumption of  complete optical pumping of the atoms into the 
$m_F=\pm4$ levels (I$_s$ = 1.1 mW/cm$^2$), while the dashed line corresponds to linear polarization with the assumption of uniform distribution among the Zeeman sublevels (I$_s$ = 1.6 mW/cm$^2$). 

\begin{figure*}[!htb]
  \includegraphics[width=15 cm]{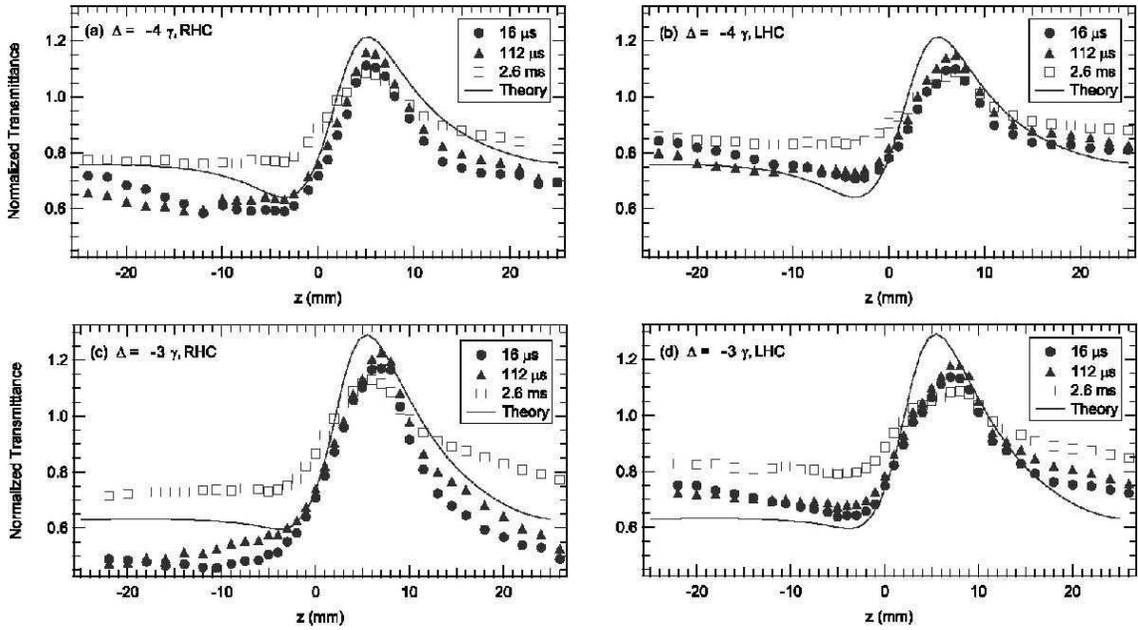}
  \caption{Measured Z-scan of a Cs cloud at $\lambda=852$ nm at different time delays with detunings $\Delta=-4\gamma$  (a,b) and $\Delta=-3\gamma$ (c,d) with respect to the $F=4\to F'=5$ transition. The data marked by open circles, triangles and crosses show the transmittance at t = 16 $\mu$s, 112 $\mu$s and 2.6 ms after the probe beam was turned on respectively. The MOT parameters  used in the calculation were $2w_z=7.2~\rm mm$,  $a_z=0.32$, $b_z=0.27$, $c_z=0.26$ (measured from camera images), and $n_{a0}=1.0\times 10^{10}~\rm cm^{-3}$(fitting parameter).}
\label{fig.Red}
\end{figure*}

The best fit to the data implies  a peak  atomic density of 
$n_{a0} = 2.3\times 10^{10}$ cm$^{-3}.$ This is lower than the measurement based on the optical pumping method, and slightly higher than the measurement using the fluorescence signal.
  As can be seen from the figure, the numerical result for a circularly polarized beam agrees reasonably with the data, although the width of the calculated scan is about 15\% narrower than the data. 
The agreement between  calculation and experiment for linear probe polarization is slightly worse.

\section{Experimental results}
\label{sec4}

\subsection{Z-Scan measurements}

\begin{figure*}[!ht]
  \includegraphics[width=15 cm]{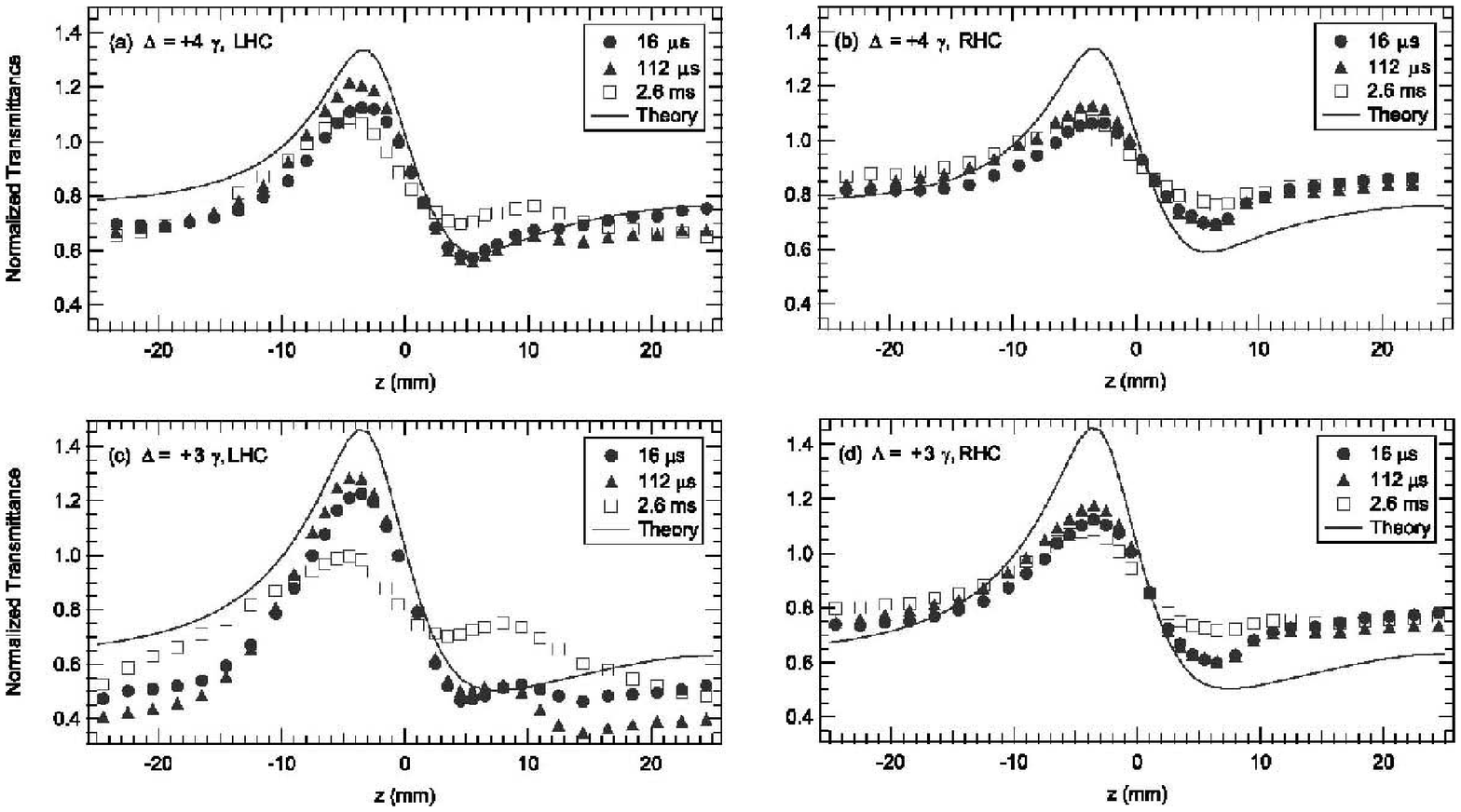}
  \caption{Measured Z-scan of a Cs cloud  at different time delays with detunings $\Delta=+4\gamma$  (a,b) and $\Delta=+3\gamma$ (c,d) with respect to the $F=4\to F'=5$ transition. The data marked by open circles, triangles and crosses show the transmittance at t = 16 $\mu$s, 112 $\mu$s and 2.6 ms after the probe beam was turned on respectively. 
The MOT parameters used in the calculation $2w_z=7.4~\rm mm$,  $a_z=-0.38$, $b_z=2.2$, $c_z=-1.1$ (measured from camera images), and $n_{a0}=0.9\times 10^{10}~\rm cm^{-3}$ (fitting parameter).}
\label{fig.Blue}
\end{figure*}

Figures \ref{fig.Red} and \ref{fig.Blue} show Z-scans measured for self-defocusing and self-focusing nonlinearities. Measurements were taken with right hand circular (RHC) and left hand circular (LHC)  probe beam polarizations at each detuning.   In all the experiments, the peak intensity of the probe beam was 1.07 W/cm$^2$.  The solid lines show the numerically calculated curves which assumed complete optical pumping of the atoms into the $m_F=\pm4$ levels as in the transmission measurements described above.

Concentrating first on the self-defocusing case with RHC polarization shown in Fig. \ref{fig.Red} a,c we see that at short times after turning off the MOT the shape of the curves is similar to that calculated numerically, while the peak density which was used as a fitting parameter was about half as large as that measured using the methods described in Sec. \ref{sec3}. We believe that the discrepancy is  due to partial atomic polarization as discussed below.
The curves have a pronounced peak when the probe waist is positioned past the center of the cloud at $z\sim 7~\rm mm$ and only a small dip, which almost disappears as the detuning is decreased from $-4\gamma$ to $-3\gamma$. The departure from the traditional S shaped form that is obtained for a pure Kerr nonlinearity\cite{Bahae89},
is due to the nonlinear absorption. 
The measurements were taken in a regime of strong saturation
at the probe beam waist, 
$s_{\Delta}=26.4$ at $\Delta=-3\gamma$  and $s_{\Delta}=15.0$ at $\Delta=-4\gamma$ where the saturation parameter is $s_{\Delta}({\bf r})=I({\bf r})/I_{s\Delta}$ and $I_{s\Delta}=I_s(1+4\Delta^2/\gamma^2)$ is the saturation intensity at finite detuning\cite{Boyd}. As we move to larger detuning the saturation parameter decreases, so the effects of nonlinear absorption are diminished and the Z-scan curves tend towards the normal symmetric S shape. 

At long times of $t=2.6 ~\rm ms$ the MOT cloud has partially dispersed and the Z-scan curve is flatter and broader with an almost complete disappearance of the valley at negative $z.$ In the intermediate regime 
of $t=112~\mu\rm s$ the peak at $z=7~\rm mm$ increases by 5-10\% compared to the value at $t=16~\mu \rm s.$ This increase is consistent with focusing, and a local increase in density, due to the radiation force 
\begin{equation}
\label{eqForce}
\mathbf{F}_{sp}({\bf r})=\hbar\,\mathbf{k}\,\frac{\gamma}{2}\frac{s({\mathbf r})}{1+4(\Delta-\mathbf{k}\cdot\mathbf{v({\bf r})})^2/\gamma^2+s({\bf r})},
\end{equation}
where  $s({\bf r})=I({\bf r})/I_s$ and 
$\mathbf{v({\bf r})}$ is the atomic velocity. As shown in Ref. \cite{Davidson99} a radiation force that decreases with distance into the MOT due to absorption of  the pushing beam results in focusing of the atoms after a finite time. Near the focal plane of the probe beam where the saturation parameter is large the peak light induced acceleration is 
$a_{\rm max}= \hbar k\gamma/(2 m)$, with $m$ the atomic mass. The time to be pushed a distance $\delta z$  is thus $\delta t=\sqrt{2\delta z/a_{\rm max}.}$
which for $^{133}$Cs evaluates to $\delta t \simeq 130 ~\mu\rm s$ for $\delta z = 0.5~\rm mm.$ This is consistent with the observation of a
larger Z-scan signal at a probe interaction time of $t=112~\mu\rm s.$

Turning now to the self-focusing case with LHC polarization shown in Fig. \ref{fig.Blue} a,c we see similar results as in the self-defocusing case at  $t=16~\mu\rm s$ except that the transmittance  
peak now appears for $z$ negative as expected. The valley in the transmittance is also more pronounced than for self-defocusing particularly at the larger detuning of $\Delta=+4\gamma.$ At the 
intermediate time of $t=112~\mu\rm s$ we see an enhanced peak which again is consistent with focusing due to the radiation pressure force.

However at the longest time of $t=2.6~\rm ms$ we see the overall flattening of the Z-scan curve together with an unexpected 
secondary peak in the transmission that appears for $z$ positive. 
This secondary peak can be qualitatively explained using Eq. (\ref{eqForce}). The radiation pressure force accelerates the atoms and Doppler shifts them to the red. Although the initial detuning is positive for self-focusing, after some time a fraction of the atoms will be Doppler shifted to  an effective negative detuning and will act to defocus the light which gives a transmission peak at positive $z.$ 
We can estimate the time for this to occur by noting that the peak time dependent Doppler shift is  
$kv(t)=k a_{\rm max} t.$ A Doppler shift of $\gamma$ is reached at $t_\gamma=2m/(\hbar k^2)$ which evaluates to $t_\gamma=~78~\mu \rm s$ for $^{133}$Cs. Thus 
for $\Delta=3\gamma$ and  
$t > t_\gamma$ we expect to see some evidence of self-defocusing, which is consistent with  the experimental data in Fig. \ref{fig.Blue} a,c at $t=2.6 ~\rm ms$. The effect is smaller when we go to $\Delta=+4\gamma$ since the scattering force is weaker and the Doppler shift must be larger in order to change the sign of the detuning. Note that no secondary features appear in Fig. \ref{fig.Red} where we start with red detuning since the Doppler shifts only move the atoms even further out of resonance.

The measurements discussed above were repeated with opposite helicity of the probe beam as shown in the right hand columns of Figs. \ref{fig.Red} and \ref{fig.Blue}. The strength of the Z-scan signal was substantially different for the two probe beam helicities.
For example for red detuning at $\Delta=-4\gamma$ the peak to valley Z-scan signal at $t=16~\mu\rm s$ from Fig. \ref{fig.Red} a,b was 0.57 for RHC polarization but only 0.39 for LHC polarization.  
For blue detuning at $\Delta=4\gamma$ the peak to valley Z-scan signal at $t=16~\mu\rm s$ from Fig. \ref{fig.Blue} a,b was 0.57 for LHC polarization but only 0.37 for RHC polarization.  
Thus for red detuning the strongest effect was obtained with RHC polarization which was the {\it opposite} polarization as the copropagating $\sigma^+$ trapping beams shown in Fig. \ref{fig.ZscanSetup}, while for blue detuning  the strongest effect  was obtained with 
LHC polarization which was the same polarization as the copropagating $\sigma^+$ trapping beams shown in Fig. \ref{fig.ZscanSetup}. We considered several possible reasons for this unexpected dependence on probe beam helicity.  All measurements were taken with the MOT magnetic field still on. However, the Zeeman shift of the cycling transition at the edge of the MOT cloud in the $x-z$ plane was less than 3 MHz which is not large enough to explain any polarization or Zeeman dependence. 

The numerical simulations were done with a two-level model which implicitly assumes complete pumping of the atomic population to the lower level of the cycling transition. Before application of the probe beam the atomic polarization will be distributed across Zeeman levels.  
The spatial localization provided by a MOT is due to the fact that $\sigma^+$ and $\sigma^-$ polarizations interact more strongly with the atoms on different sides of the cloud. This naturally leads to a spatial variation of the atomic polarization. Indeed a  rate equation model\cite{CClin02} predicts a  polarization of 20\% or more at the edges of a Rb MOT. At very short times after the probe beam is turned on, or if the optical pumping due to the probe beam were only partially successful, we would expect a polarization dependent interaction that depended on the helicity of the probe. However, Figs. \ref{fig.Red} and \ref{fig.Blue} show that  the helicity of the probe that interacts most strongly with the MOT switches when we change the sign of the detuning.
 
\begin{figure}[htb]
  \includegraphics[width=8.2cm]{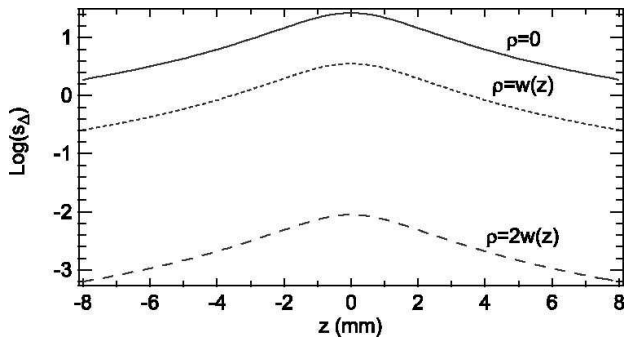}
\caption{Saturation parameter of the probe beam in the atomic cloud at different radial positions $\rho=\sqrt{x^2+y^2}$. $w(z)=w_0\sqrt{1+\lambda^2 z^2/(\pi^2 w_0^4)}$ is the z-dependent probe beam waist. }
\label{fig.sat}
\end{figure}

This effect can be explained qualitatively in the following way. 
Looking at Fig. \ref{fig.sat} we see that the effective off-resonance saturation parameter of the probe beam varies by many orders of magnitude across the atomic cloud.  The volume that determines the nonlinear diffraction of the probe beam can reasonably be taken to extend out to off-axis distances of a beam waist where the off-resonance saturation 
parameter is as small as 0.001 at the edges of the cloud for a centered probe beam. 
Let us assume the atomic population before application of the probe beam is distributed across the Zeeman levels.  Then the longest optical pumping time, that for transferring an atom in $m_F=-4 $ to $m_F=4,$ will be, neglecting the excited state branching ratios,  
$t_{\rm pump}\sim  (2/\gamma)[(1+s_\Delta)/ s_{\Delta}]\sum_{m_F=-4}^{m_F=4}1/|C_{m_F,m_F+1}|^2$ where the Clebsch-Gordan coefficients are normalized such that $C_{m_F=4,m_{F'}=5}=1$. 
We find $t_{\rm pump}\sim 20~\rm\mu s$ at $s_{\Delta}=1$ and $t_{\rm pump}\sim 1~\rm m s$ at 
$s_{\Delta}=0.01.$ The implication of this estimate is that the probe beam does not completely polarize the volume of the MOT that it interacts with,  even at times as long as several hundred $\mu\rm s$.
At the latest time shown in Figs. \ref{fig.Red} and \ref{fig.Blue},
$t=2.6~\rm ms$, the optical pumping is substantially complete, and there is only a very small difference between the data taken with opposite probe beam helicities, except for the effects of the radiation force as discussed above.

The Z-scan signal is intrinsically due to the nonlinear effects of self-focusing and self-defocusing. These effects are strongest when a circularly polarized probe interacts with a fully polarized atomic sample. However, the radiation forces that shift the detuning to the red, are also maximized for a fully polarized atomic sample. 
When the probe is blue detuned the radiation forces will initially increase the strength of the interaction so that a probe polarization that couples strongly to the atomic polarization will give a larger effect. In the opposite case of red detuning of the probe the pushing forces only serve to decrease the strength of the interaction so that the Z-scan signal will be strongest when the pushing is reduced. 
Although the MOT has opposite atomic polarization for positive and negative $z$ and is not expected to have any net polarization imbalance averaged over the cloud, the probe beam is attenuated as it propagates so that
the front edge of the cloud has a stronger impact on the Z-scan signal.
We thus expect a larger signal when a red detuned probe has a polarization that gives a weaker atomic coupling, and less pushing forces, on the front side of the cloud,  and a larger signal when a blue detuned probe has a polarization that gives a stronger atomic coupling on the front side of the cloud. 
These arguments suggest that a red-detuned probe will give a larger signal when it has the opposite helicity of the trapping beam that has the same momentum projection along $\hat z$, and  that a blue-detuned probe will give a larger signal when it has the same helicity as the trapping beam. This is indeed what is observed in Figs. \ref{fig.Red} and \ref{fig.Blue}.

As a further check on this explanation we changed the sign of the magnetic field and reversed the helicities of all the trapping laser beams. We then found that the data were the same as that measured previously provided we also flipped the helicity of the probe beam. This supports the conclusion that the dependence on probe beam polarization is due to the spatial distribution of atomic polarization inside the MOT in the presence of imperfect optical pumping and radiative forces. It is interesting to note that the number density 
deduced from  transmission scans shown in Fig. \ref{fig.Abs} would be higher, and thus in closer agreement with the other density measurements discussed in Sec. \ref{sec3}, if we took into account a partial polarization of the atoms. 
However the transmission scans,  which were taken without a pinhole, and therefore depend only on the total transmission, and not the shape of the wavefront, show no  helicity dependence. We conclude that Z-scan measurements provide a signal that is comparatively sensitive to the atomic polarization distribution. The extent to which a Z-scan could be used to measure the polarization distribution quantitatively remains an open question.

\begin{figure}[b]
  \includegraphics[width=8.2cm]{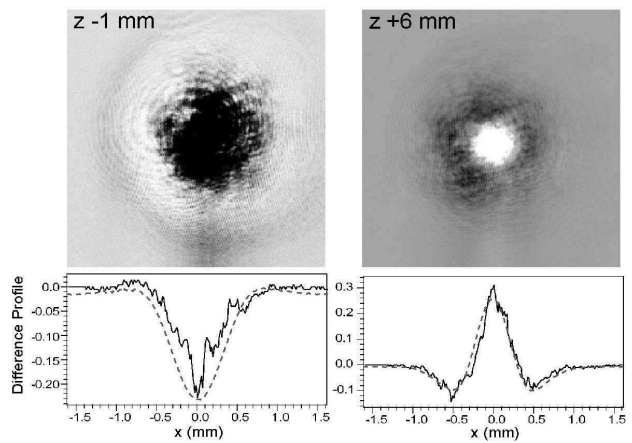}
  \caption{Transmitted probe beam intensity for $\Delta=-3\gamma.$
The pictures show the intensity minus the intensity without a MOT 
for $ z = -1 ~\rm mm$ and $ +6~\rm  mm$. The bottom figures show the line profiles across the center of the beam, which are normalized to the center intensity of the beam without cold atoms. The dashed line is the result of numerical calculations.
The peak MOT density was  $n_{a0}=0.7\times10^{10}~\rm cm^{-3}$ while $w_z,a_z,b_z,c_z$ were  the same as in Fig. \ref{fig.Blue}.}
  \label{fig.picred}
\end{figure}

\begin{figure}[!t]
  \includegraphics[width=8.2cm]{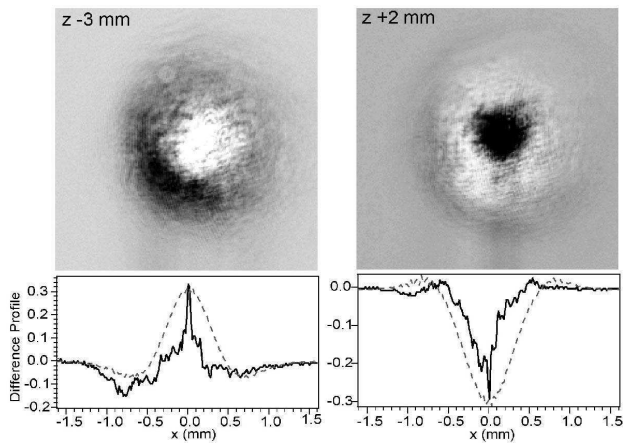}
  \caption{Transmitted probe beam intensity for $\Delta=+3\gamma$
at $ z = -3 ~\rm mm$ and $ +2~\rm  mm$.
All other parameters are the same as  in Fig. \ref{fig.picred}.
}
  \label{fig.picblue}
\end{figure}

\subsection{Spatial redistribution of intensity}

In addition to the Z-scan transmittance measurements, the transmitted probe beam far field distribution was observed directly using a CCD camera. To do so, the pinhole and photodetector $D$ in Fig. \ref{fig.ZscanSetup} were removed, and a lens ($f=100~\rm mm$) was used to image a plane at $z= 72~\rm  mm$ after the cloud center  onto the camera. Pictures were recorded with the beam waist  at different positions relative to the center of the cloud for 
self-defocusing and self-focusing cases as shown in Figs. \ref{fig.picred} and \ref{fig.picblue}. To illustrate the self-defocusing or self-focusing of the beam, the difference between the intensity distributions with and without cold atoms is shown.  In Fig. \ref{fig.picred} we see the result with the  beam waist  near the center of the cloud ($z = -1~\rm  mm$) and at the edge of the cloud close to the detector ($z = +6 ~\rm  mm$). At $z = -1~\rm  mm$, the beam gets focused faster due to the nonlinearity, so the far field transmitted beam has an additional  divergence  compared to the linear case.  Thus the central part of the picture is dark. At $z = +6 ~\rm mm,$ the beam becomes less focused due to the same effect, so that in the  far field the beam is  more converged relative to the linear case, which gives a bright region in the center of the picture. For the two pictures in Fig. \ref{fig.picblue} we have similar results with the roles of positive and negative $z$ interchanged due to the opposite sign of the nonlinearity. 

The lower parts of Figs.  \ref{fig.picred} and \ref{fig.picblue} show that numerical calculations predict the transverse beam profiles with an accuracy similar to that seen for  the Z-scan transmittance curves. 
We see that the agreement is best for a self-defocusing nonlinearity while in the self-focusing case there is a tendency towards localized maxima and minima in the transverse profiles. Similar phenomena were reported in Ref. \cite{kaiser}.  Additional numerical calculations at 5-10 times higher densities reveal strong filamentation of the transmitted beam. Future work will investigate propagation effects in this regime experimentally.

\section{Discussion}
We have used Z-scan measurements to characterize the nonlinear optical response of a cold atomic cloud.  We show that the measured data agree with a two-level atomic model at short times provided the probe helicity is chosen to match the helicity of the trapping beams 
on the front side of the cloud.  At longer times modifications to the Z-scan occur because of radiation pressure forces. 
This results in some additional focusing of the cloud and the appearance of a secondary Z-scan peak when a blue detuned probe  accelerates the atoms past resonance, to give a partially red detuned response. Transverse profiles of the probe beam show focusing and defocusing features consistent with the Z-scan transmittance measurements. Future work will study clouds with larger optical thickness where we expect modulational instabilities of the probe beam
to lead to small scale modifications of the atomic distribution.

\begin{acknowledgments}
We thank Thad Walker for helpful discussions.
Support was provided by The University of Wisconsin Graduate School, the Alfred P. Sloan foundation, and NSF grant  PHY-0210357.  
\end{acknowledgments}

\bibliography{zscansubmit}

\end{document}